\documentclass[%
 aip,
 amsmath,amssymb,
 reprint,%
]{revtex4-1}

\usepackage{graphicx}
\usepackage{dcolumn}
\usepackage{bm}

\usepackage[utf8]{inputenc}
\usepackage[T1]{fontenc}
\usepackage{mathptmx}
\usepackage{etoolbox}

\usepackage[normalem]{ulem}

\makeatletter
\def\@email#1#2{%
 \endgroup
 \patchcmd{\titleblock@produce}
  {\frontmatter@RRAPformat}
  {\frontmatter@RRAPformat{\produce@RRAP{*#1\href{mailto:#2}{#2}}}\frontmatter@RRAPformat}
  {}{}
}%
\makeatother

\usepackage{amsthm}
\theoremstyle{definition}

\usepackage{xcolor}

\begin{document}

\preprint{AIP/123-QED}

\title[Magnetic Reservoir Computing]{A perspective on physical reservoir computing with nanomagnetic devices}
\author{Dan A Allwood}
 \affiliation{Department of Materials Science and Engineering,  University of Sheffield,  S1 3JD,  United Kingdom}
\author{Matthew O A Ellis}%
\affiliation{ 
Department of Computer Science, University of Sheffield, S1 4DP, United Kingdom
}%

\author{David Griffin}
\affiliation{%
Department of Computer Science, University of York, YO10 5GH, United Kingdom
}%

\author{Thomas J Hayward}
 \email{t.hayward@sheffield.ac.uk}
 \affiliation{Department of Materials Science and Engineering, University of Sheffield, S1 3JD,  United Kingdom}

\author{Luca Manneschi}%
\affiliation{ 
Department of Computer Science,  University of Sheffield, S1 4DP, United Kingdom
}%

 \author{Mohammad F KH Musameh}
 \affiliation{Department of Electronic Engineering, University of York, YO10 5DD, United Kingdom}
 
\author{Simon O'Keefe}
\affiliation{Department of Computer Science, University of York, YO10 5GH, United Kingdom}
 
\author{Susan Stepney}
\affiliation{Department of Computer Science, University of York, YO10 5GH, United Kingdom}
 
\author{Charles Swindells}
 \affiliation{Department of Materials Science and Engineering, University of Sheffield, S1 3JD, United Kingdom}

 \author{Martin A Trefzer}
 \affiliation{Department of Electronic Engineering,  University of York, YO10 5DD, United Kingdom}
 
\author{Eleni Vasilaki}
\email{e.vasilaki@sheffield.ac.uk}
\affiliation{ 
Department of Computer Science,  University of Sheffield, S1 4DP, United Kingdom
}%
  
\author{Guru Venkat}
 \affiliation{Department of Materials Science and Engineering, University of Sheffield, S1 3JD,  United Kingdom}
 
\author{Ian Vidamour}
  \affiliation{Department of Materials Science and Engineering, University of Sheffield, S1 3JD,  United Kingdom}
  \affiliation{ 
Department of Computer Science,  University of Sheffield, S1 4DP, United Kingdom
}%
 
\author{Chester Wringe}
\affiliation{Department of Computer Science, University of York, YO10 5GH, United Kingdom}

\date{\today}

\begin{abstract}
Neural networks have revolutionized the area of artificial intelligence and introduced transformative applications to almost every scientific field and industry. However, this success comes at a great price; the energy requirements for training advanced models are unsustainable. One promising way to address this pressing issue is by developing low-energy neuromorphic hardware that directly supports the algorithm's requirements. The intrinsic non-volatility, non-linearity, and memory of spintronic devices make them appealing candidates for neuromorphic devices. Here we focus on the reservoir computing paradigm, a recurrent network with a simple training algorithm suitable for computation with spintronic devices since they can provide the properties of non-linearity and memory. We review technologies and methods for developing neuromorphic spintronic devices and conclude with critical open issues to address before such devices become widely used. 
\end{abstract}

\maketitle

\section{\label{sec:intro}Introduction}

Neural networks are widely used across various sectors to perform challenging data analysis tasks, but the high energy cost of training increasingly complex models is an escalating problem. More specifically, for training a state-of-the-art model, a Transformer with 213M parameters, the CO2 emissions were 626,155 lbs (including neural architecture search), while driving a car (average fuel consumption),  for one lifetime, the Co2 emissions were only 126,000 lbs \cite{strubell2019energy}. One solution to the energy issue is to create new hardware platforms for neuromorphic computation using functional materials that intrinsically perform the required computation, potentially achieving greater efficiency than conventional CMOS approaches that merely simulate these. 
Recurrent neural networks (RNNs) are inspired by the high interconnectivity of biological systems and are a potent tool for tasks involving complex temporal data sequences. However, their temporal interconnectivity requires complex training methods. Such methods are computationally expensive and challenging to implement on hardware. The reservoir computing (RC) paradigm provides a solution using an RNN with fixed, random synaptic weights (the reservoir) that transforms inputs into higher dimensional representations before passing them to a single feed-forward output layer. The weights of this output layer can be calculated by minimizing an error function defined, for instance, as the squared difference between the desired and the predicted output. The output layer contains no temporal dependencies, and thus training becomes relatively trivial. Ultimately, the reservoir does not need to be a neural network; it can be any suitable nonlinear system that exhibits hysteresis. RC is particularly well suited to neuromorphic hardware-based implementations. Since the learning process does not interfere with the reservoir dynamics, we may use any material device which provides appropriately complex dynamics and memory in the place of a neural network reservoir.

\begin{figure}
    \centering
    \includegraphics[width=\columnwidth]{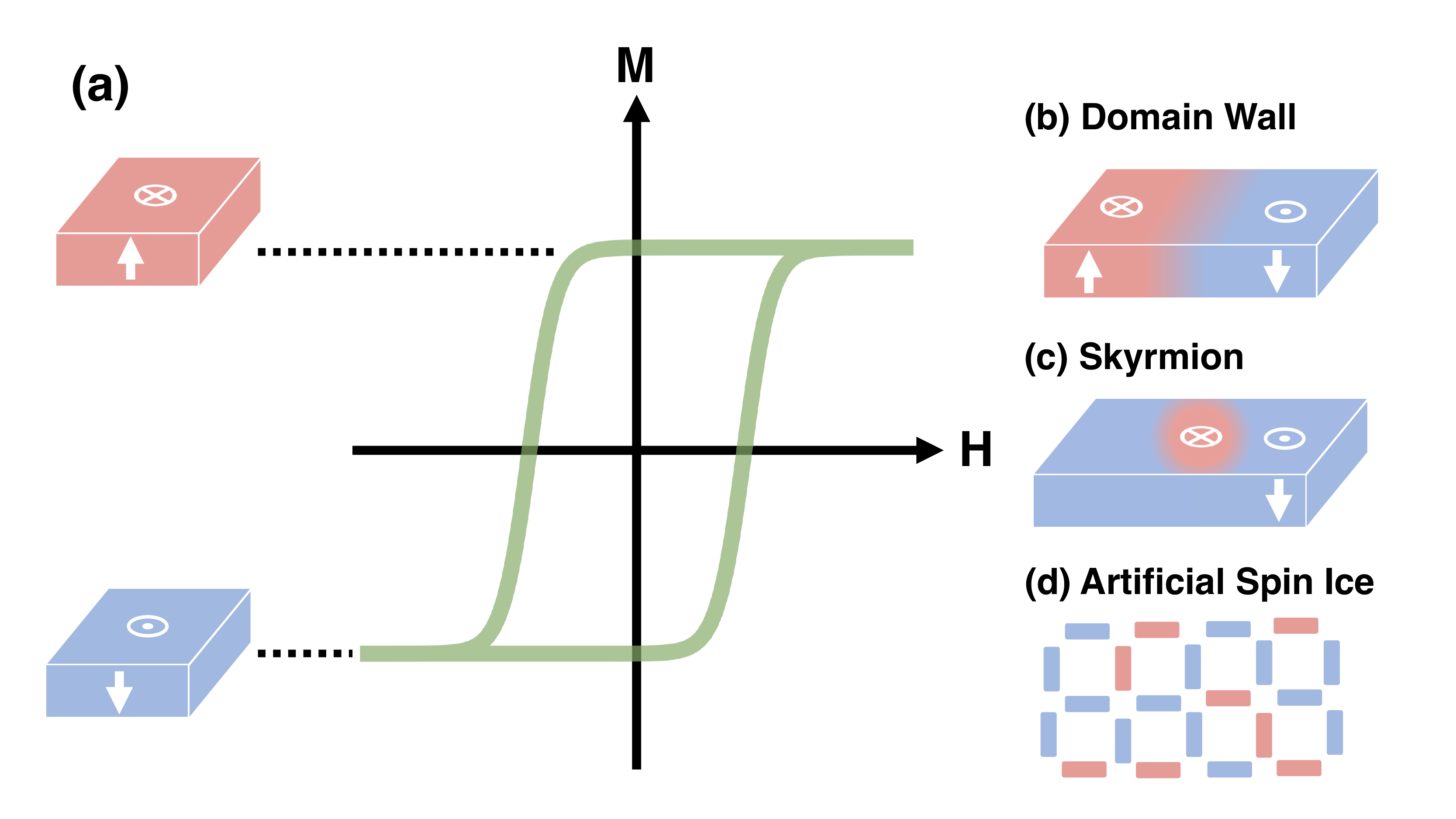}
    \caption{Magnetic hysteresis loop and complex magnetic textures. (a) Bistable hysteresis loop where the applied magnetic field (H) controls the orientation of the magnetization (M). (b-d) Examples of complex magnetic textures are domain walls, skyrmion, and artificial spin ice. }
    \label{fig:hysteresis}
\end{figure}

There are explorative reservoirs from different technologies, including photonic\cite{nakajima2021scalable}, mechanical\cite{dion2018reservoir}, and memristive\cite{gaurav2022reservoir} systems. Nanomagnetic systems have properties that make them particularly well-suited to act as reservoirs. For example, the magnetic hysteresis loop depicted in Figure \ref{fig:hysteresis} shows a non-linear response (the net magnetization) to a stimulus (the applied field). Bistable remnant magnetization states, shown schematically in Figure \ref{fig:hysteresis} (a), can be the basis for the system's memory. Furthermore, in extended systems, interactions between moments give rise to a wealth of magnetization textures with complex dynamics that provide a rich playground to explore novel devices. Some example textures are shown in Figure \ref{fig:hysteresis} (b-d), showing magnetic domain wall, skyrmion, and artificial spin ice systems. Short-range exchange and longer-range magnetostatic interactions offer {\it in materia} pathways to creating reservoirs with multiple physical nodes without the need for complex material synapses between nodes. The historical use of magnetic materials in hard-disk drives, sensors, and random access memories means integration with CMOS and techniques for reading (e.g., magnetoresistance effects) and writing data (e.g., magnetic fields, spin torque effects) are also well-established.  

In this perspective, we will first review the approaches for creating {\it in materia} reservoirs using nanomagnetic materials and their various strengths and weaknesses. We will discuss the most common training methods that map their physical behaviors into meaningful data outputs. Next, we will discuss simulation tools that can assist in exploring the feasibility of reservoir computing with different magnetic systems and the characterization methods and benchmark problems commonly used to establish computational capability. Finally, we present some key challenges in the field and potential approaches to address these.

\section{Materials and devices}
Due to their attractive properties, several nanomagnetic systems have been deemed suitable as reservoirs. These systems include: spin torque oscillators (STOs)\cite{Torrejon++2017,riou__2019,Markovic++2019,furuta_2018,jiang_2019}; spin ice arrays\cite{Jensen++2018:spinice,Jensen++2020:spinice,zhou2020reservoir,hon_2021,Nomura_2019, https://doi.org/10.48550/arxiv.2107.08941}; skyrmion textures\cite{Pinna++2020,prychynenko__2018,jiang_2019}; superparamagnetic arrays\cite{Welbourne++2021}; magnonic systems\cite{watt2020reservoir} and domain wall devices\cite{Dawidek++2021,ababei__2021}. Most studies are in simulations, although some demonstrations of RC with real devices have been performed, providing important evidence of real-world feasibility\cite{Torrejon++2017,https://doi.org/10.48550/arxiv.2107.08941,watt2020reservoir}. 

\begin{figure}[htb]
    \centering
    \includegraphics[width=\columnwidth]{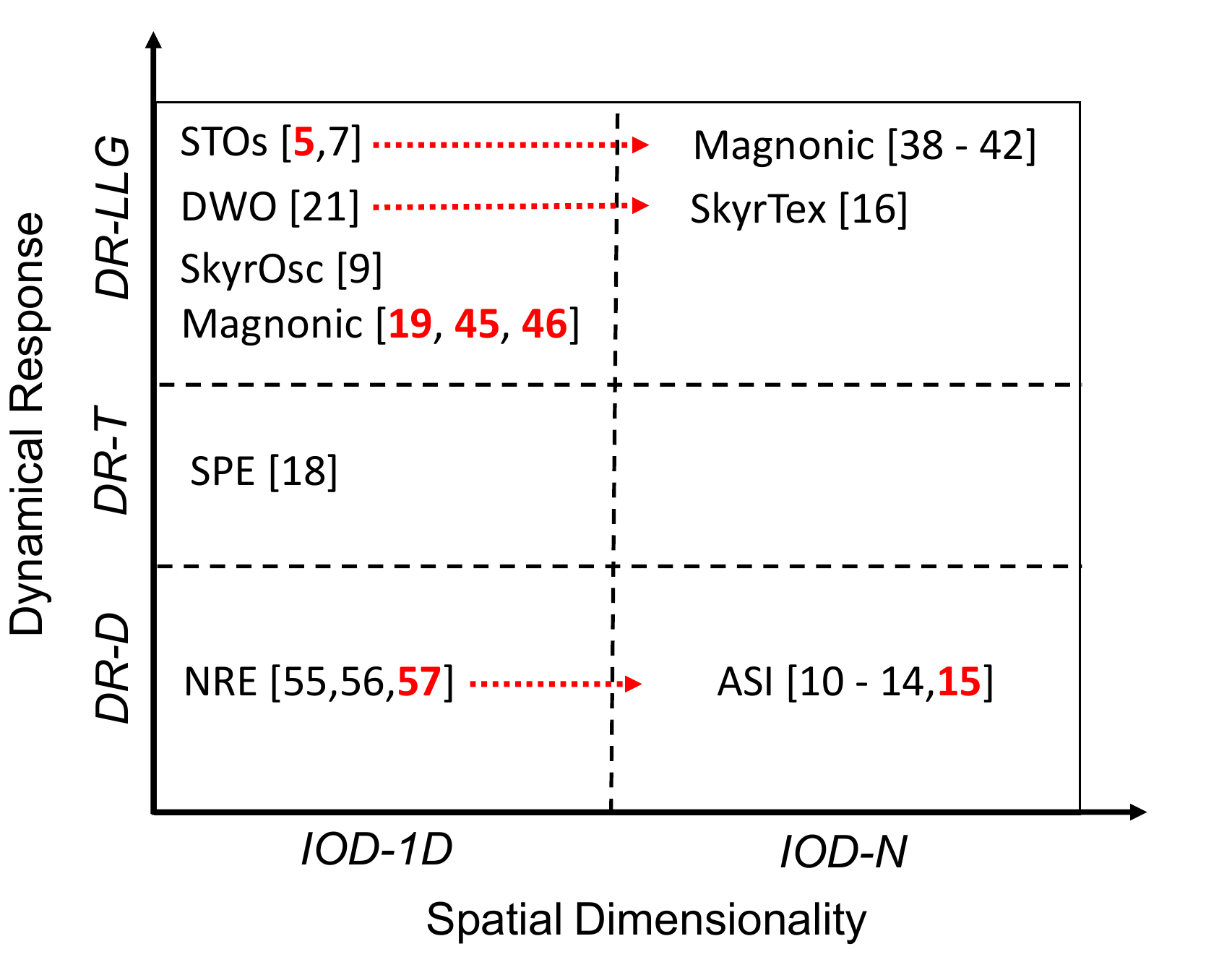}
    \caption{Classification of magnetic reservoir proposals by Input/Output Dimensionality (IOD) and Dynamical Response (DR). IOD: IOD-1D = single dynamical node, IOD-N = multiple spatial dimensions. DR: DR-D = driven by external clock stimulus, DR-T = dynamics governed by thermal activation/relaxation, DR-LLG = dynamics governed by LLG equation. References in bold red text are experimental demonstrations; all other references are simulation-based demonstrations. Red arrows represent systems where the state-of-the-art is an IOD-1D demonstration, but there are clear {\it in materia} approaches available to create IOD-N reservoirs. Key: STOs = spin torque oscillators, DWO = domain wall oscillators, SkyrOsc = skyrmion oscillator, SPE = super-paramagnet ensemble, NRE = nanoring ensemble, Magnonic = magnonic reservoir, SkyTex = skyrmion texture, ASI = artificial spin ice. }
    \label{fig:taxonomy}
\end{figure}

In general, nanomagnetic reservoirs can be classified based on several characteristics (e.g., energy consumption, operating speed, and device size). Here we introduce a taxonomy that classifies proposed devices by (a) Input/Output Dimensionality (IOD) and (b) Dynamical Response (DR) (Figure ~\ref{fig:taxonomy}). 

For IOD, RC requires multiple outputs from the reservoir (i.e., simultaneous measures of reservoir state) and benefits from multiple, simultaneous data inputs. Many devices proposed for use in RC are simple dynamical nodes with only a single input and output (IOD-1D). To use IOD-1D as reservoirs, we must expand the dimensionality of input and output data by using time-multiplexing techniques \cite{appeltant_2011}, an approach often referred to as "delay line" RC. However, other proposed devices consist of many spatially distributed, interacting elements/regions. These naturally possess \emph{N} dimensional state vectors and thus offer an \emph{in materia} pathway to defining multiple input and output dimensions (IOD-N). Reservoirs containing multiple \emph{non-interacting} devices can also be powerful, providing that each device offers a different non-linear mapping of input signals\cite{furuta_2018}.

For DR, many proposed magnetic reservoirs exploit the damped, oscillatory motion of individual magnetic moments, as described by the Landau-Lifshitz-Gilbert (LLG) equation of motion (DR-LLG). These dynamics have high MHz-THz frequencies and ns decay times for ferromagnetic materials, making them well-suited to high-speed data processing applications. RC is also ideal for real-time signal processing where reservoir timescales must match external signals with low or high frequencies. However, as the dynamics of DR-LLG systems occur on nanosecond timescales, they are too fast for many real-time tasks; one must use external electronics to "speed-up" data input or improve long-term dependencies via delay lines. Effectively, we treat the magnetic devices as non-linear activation functions with short-term temporal dependencies\cite{riou__2019}.

Other magnetic devices do not naturally relax their state without applied stimuli; external clocking stimuli determine the timescales of these dynamically driven (DR-D) systems. By choosing the clock frequency, these systems can operate at any timescale \emph{longer} than intrinsic magnetization dynamics. Therefore, they are naturally well-suited to real-time data analysis but may be less energy efficient than DR-LLG devices. A final class of reservoirs directly exploits thermally activated magnetization dynamics to provide transitions between magnetic states (DR-T). These are interesting as they directly exploit aggregated thermal effects to increase energy efficiency, whereas, in most device proposals, thermal effects introduce stochasticity, reducing performance in computational tasks. Furthermore, as the timescales of thermal activation can be changed dramatically (down to \textasciitilde10s of nanoseconds\cite{thermalrelaxns}) by changing the size of the systems' energy barriers, it should be possible to tune these systems dynamics to be compatible with a variety of real-time tasks. However, their stability to variations in operating temperature requires careful exploration.   

In the following section, we briefly review the wide range of device proposals within this framework and discuss their other potential merits and limitations.

\subsection{Nanomagnetic Oscillators}

Spin torque oscillators\cite{7505988} (STOs) (IOD-1D, DR-LLG) use the same magnetic tunnel junction (MTJ) technology that forms the basis of contemporary MRAM devices\cite{kent_worledge_2015}. At the most basic level, MTJs consist of two thin ferromagnetic layers separated by a thin insulating barrier in a “spin valve” configuration. One of the ferromagnetic layers is free to change its magnetization direction (free layer). The other is “pinned” into a fixed state (pinned layer) by an adjacent antiferromagnetic layer. Passing a DC electrical current through the multilayer excites oscillation of the magnetization direction of the free layer due to spin torque effects\cite{slonczewski_1996,berger_1996}, with frequencies in the range 100s of MHz to 10s of GHz, depending on the details of the oscillator’s design and stimuli applied to it. When the free layer magnetization oscillates, it produces oscillations in the electrical resistance of the MTJ via the tunnel magnetoresistance (TMR) effect. TMR can be detected as voltage signals with amplitudes as large as 10s of mV\cite{tsunegi__2016}. The amplitude of STO oscillations varies non-linearly with current and typically decays over timescales of $\sim$100s of ns\cite{Torrejon++2017}.

Torrejon \emph{et al.}\cite{Torrejon++2017} demonstrated RC experimentally using a single sub-micrometer STO device using the time-multiplexed approach of Appletant \emph{et al.}\cite{appeltant_2011} Input signals are given to the STO by modulating the amplitude of the DC driving current, with the readout being the power output of the STO. Using this approach, the authors achieved state-of-the-art performance when classifying spoken digits from the TI-46 database\cite{TI-46}. Alternative input and output approaches (e.g., frequency modulated input, phase modulated output) can also create richer reservoir transformations and improve performance in tasks\cite{Markovic++2019}.

STOs have many attractive properties. Foremost among these is that MTJs are a well-established commercial technology and are fully compatible with conventional CMOS platforms, providing a clear path to the realization of devices. Furthermore, they can be scaled down substantially from the sub-micrometer dimensions studied by Torrejon \emph{et al.} to \textasciitilde10 nm, creating device designs that are both dense and energy efficient (\textasciitilde1 µW per STO). 

While recent demonstrations have focused on time-multiplexed RC schemes, interconnections between STOs allow them to couple to each other\cite{slavin_2009,houshang_2015}, potentially facilitating \emph{N}-dimensional reservoirs. Current approaches to neuromorphic computation with STOs have used external electrical interconnects to achieve this\cite{romera_2018}. Still, STOs can interact/synchronize via magnetic interactions \cite{houshang_2015,zahedinejad_2019}, allowing for simpler and more elegant device designs. 

Other types of magnetic oscillators can also be used as reservoirs. Ababei \emph{et al.} used simulations to show that a single magnetic domain wall (DW) oscillating within a geometrically defined potential well in a nickel nanowire can create a reservoir capable of classifying a variety of different signals\cite{ababei__2021} (IOD-1D, DR-LLG). In this approach, the DW's dynamics are dictated by device geometry and, therefore, should be highly tunable. Furthermore, DWs naturally produce monopole-like magnetic fields\cite{PhysRevB.81.020410,hayward_2010}, allowing inter-device interactions to expand reservoir dimensionality. In a similar modeling study, Jiang \emph{et al.} use the dynamics of a single magnetic skyrmion (i.e., a topologically protected "bubble" of non-uniform magnetization) within a geometrically-defined potential to make an effective reservoir\cite{jiang_2019} (IOD-1D, DR-LLG). 

\subsection{Magnonic Systems}

When driven at microwave frequencies, magnetic materials exhibit phase-coherent collective excitations known as spin waves (SWs), the quasiparticle of which is the magnon. The frequencies of  SWs depend strongly on both material properties and induced magnetic anisotropies imposed by the system's geometry. The magnetic damping parameter of a material quantifies how efficiently SWs dissipate into the lattice and must be minimized by using materials such as permalloy (NiFe) or Yttrium Iron Garnet (YIG)\cite{haldar2021functional} to reduce losses. Boundaries and interfaces within a material allow for complex SW interference patterns to form, akin to reservoir work involving the pattern of water waves in a bucket\cite{fernando2003pattern}. This high degree of tunability provides a rich parameter space for useful computation. At the same time, the intrinsic spatial variation of interference effects makes spin waves an ideal phenomenon for developing IOD-N reservoirs. As these approaches directly exploit magnetization dynamics, they all have class DR-LLG.

Papp \emph{et al.}\cite{papp2021characterization} used micromagnetic simulations to characterize the computational potential of a simulated SW reservoir based on a film of YIG (IOD-N, DR-LLG) using task agnostic metrics. Modulating an RF excitation from a waveguide on one side of the film provides the input. The output is the time-averaged signal response at points across the system. Patterned dots of material with perpendicular magnetic anisotropy (PMA) on the surface of the YIG provided a non-uniform magnetic field, which locally altered the SW dispersion, resulting in a non-linear response. The system's response strongly depends on the regime at which the SWs were driven. For example, too high an input excitation would drive the system toward chaos. Nakane \emph{et al.} suggest that magnetoelastic effects in multiferroic systems could provide energy-efficient excitation of spin wave reservoirs \cite{Nakane++2018,Nakane++2019,Nakane++2021}.

In another simulation-based study, Dale \emph{et al.} explored the limits of magnonic RC \cite{dale2021computing} by considering thin films of Co, Fe, and Ni with \textasciitilde100 nm lateral dimensions (IOD-N, DR-LLG). These were split into a regular grid of up to 900 5 nm x 5 nm nodes which were excited with local magnetic fields for data input and with the local 3d magnetization state of each node providing output. SWs reflect from edges forming interference patterns that provide a complex, transient transformation of input data. For larger numbers of nodes at 0~K, the system achieves impressive task-agnostic metric scores (see section \ref{CHARC}) and an error of about 1\% for a NARMA-30 task. As expected, the introduction of temperature to the simulation drastically reduced performance. Experimental realization of an equivalent device would be highly challenging, and cooling devices to cryogenic temperatures are unlikely to be energy efficient. Hence, further work is required to explore device designs that are feasible to fabricate and robust to higher temperatures.

Physical devices based on SWs are challenging to realize, partly due to devices operating at non-zero temperatures, which can alter magnon behaviour\cite{agrawal2013direct}. Watt \emph{et al.} experimentally demonstrated an SW-based system with a time-multiplexed active ring resonator approach \cite{watt2020reservoir,watt2021implementing,watt2021enhancing} (IOD-1D, DR-LLG). The system consists of two antennas on each side of a strip of YIG: one to excite SWs and the other to detect them. The amplified microwave output signal is fed back into the input antenna to shift the phase of the frequencies within the YIG. An increase in gain stabilizes the SWs, until the threshold at which chaotic behavior occurs. This time-delayed transition to a steady-state condition acts as a fading memory within the system without needing external time-delayed input\cite{watt2021implementing}. 

Magnonic systems provide a potential platform for fast, low-power reservoir computing. However, they require high-quality growth of insulating magnetic films such as YIG and may show the best performance at low temperatures. Further work, particularly on experimental SW-based devices, is needed to explore their potential fully.

\subsection{Artifical Spin Ice Systems}

Artificial spin ice (ASI) arrays consist of magnetically-bistable nanoscale islands of soft magnetic materials (e.g. permalloy) arranged into tightly spaced, periodic lattices of various geometries\cite{skjaervo_2019}. Magnetostatic fields created by the elements in these lattices mean that any given nanomagnet's free energy depends strongly on its magnetization's direction relative to its neighbors. Thus, the physics of ASIs are emergent, with complex collective behaviors deriving from simple interactions at an array's vertices. They provide a rich playground to explore various physical phenomena, including phase transitions, emergent magnetic monopoles, and magnetic frustration. Dynamics in these experiments are typically driven by applying external magnetic fields or directly heating the arrays. Studies have explored a wide range of geometries, including, for example, square lattices\cite{wang_2007}, kagome lattices\cite{wills_2002}, and pinwheel lattices\cite{gliga_2017}. Fully-connected ASIs can also be created where exchange interactions mediate interactions between vertices, and switching occurs by the propagation of DWs\cite{mellado_2010}. 

ASIs are particularly effective systems for RC. They consist of large numbers of spatially distributed elements that interact strongly with their neighbors without the need for layers of interconnects, offering a natural platform for realizing IOD-N reservoirs. Their complex and highly tuneable dynamics (e.g., via their large geometric phase space) promise a wealth of non-linear transforms of input data. Their dynamics are typically "clocked" by external stimuli, making them examples of DR-D systems. 

Initial simulation-based studies by Jensen \emph{et. al.}  show that the large binary state space of ASIs can be fully exploited computationally\cite{Jensen++2018:spinice} and that even subsampled representations of the magnetic state retain substantial computational power when used as outputs\cite{Jensen++2020:spinice} (IOD-N, DR-D). Other simulation studies have demonstrated that data can be input using the configurations of individual, or small groups, of islands\cite{zhou2020reservoir,hon_2021,Nomura_2019}. These studies provide strong evidence that the large numbers of interacting, binary degrees of freedom in ASIs are a genuine asset for creating IOD-N RC platforms.

There are substantial challenges to experimentally demonstrating the computational abilities of ASIs. While it is possible to envision ASIs constructed from dense arrays of individually addressable MTJs that would facilitate data input and output, the fabrication of such devices is beyond what is achievable in most research laboratories. Thus, alternative methods must be used to determine how the microstates of ASIs vary when subjected to complex field sequences. Gartside \emph{et al.} have used ferromagnetic resonance measurements to "fingerprint" the microstates of an ASI\cite{https://doi.org/10.48550/arxiv.2107.08941}. Their novel approach led to the first experimental demonstration of RC using an ASI to perform signal reconstruction and time series prediction tasks (IOD-N, DR-D). Globally applied magnetic fields were used to "clock" the ASI-based reservoirs. Still, such fields would likely be energy intensive for device-level implementations, and alternative clocking methods, e.g., spin or spin-orbit torque effects, will be necessary. 

While the potential strength of ASIs as reservoirs stems from interactions between elements, Welbourne \emph{et al.} have shown that collections of magnetic islands are capable of computation even in the non-interacting limit~\citep{Welbourne++2021}. In a simulation study, the authors used ensembles of voltage-controlled super-paramagnetic islands as time-multiplexed reservoirs, demonstrating high performance in both chaotic series prediction and spoken digit recognition tasks (IOD-1D, DR-T). Energy consumption was estimated to be \textasciitilde24 fJ per input, which makes the proposed devices attractive for edge computing applications where low power consumption is vital. However, RC systems contain multiple components beyond the reservoir material itself. Further research is needed to understand how the total power consumption is related to that of the reservoir itself.

\subsection{Skyrmion and Domain Wall Ensembles}
Magnetic nanostructures can support a variety of stable, non-uniform magnetization textures. Examples of such textures are domain walls and magnetic skyrmions that exhibit complex dynamics and strong interactions when placed in close proximity.

Skyrmions are topologically-protected bubble-like magnetization textures stabilized in magnetic materials that exhibit strong Dzyaloshinskii-Moriya interactions\cite{fert_2017}. These can be found in single crystal bulk magnetic materials with non-centrosymmetric lattices (e.g., MnSi\cite{muhlbau_2009}) or in thin film systems that lack inversion symmetry (e.g., Pt/Co/Ir multilayers\cite{moreau_2016}). Skyrmions can be displaced at relatively low current densities using spin-orbit torques and produce unique electrical signatures via the topological Hall effect\cite{fert_2017}. In extended systems, skyrmion textures/fabrics can be formed; these interpolate between particle-like individual skyrmions and complex domain structures bounded by chiral domain walls.

Pinna \emph{et al.} have studied the feasibility of reservoir computing with skyrmion textures using micromagnetic simulations\cite{Pinna++2020} (IOD-N, DR-LLG). These were excited using spin torque effects by passing current between two electrical contacts. The readout could be either (i) a time-multiplexed sampling of the device's anisotropic magnetoresistance or (ii) multiple spatially-resolved samples of the textures' magnetization configurations. 
The authors showed that the device could classify sine and square waves within random sequences provided that the dynamics of the input signals were well-matched to those of the skyrmions dynamics, which were in the GHz regime. However, there are a variety of hurdles still to be overcome for experimental realizations. Chief amongst these is that 
for temperatures above T = 100 K, thermal noise obscures the
the anisotropic magnetoresistance (AMR) signals\cite{Pinna++2020}, indicating a need for alternative readout mechanisms.

Dawidek \emph{et al.} have proposed an alternative reservoir design that exploits stochastic interactions between domain walls in a patterned array of interconnected, micron-scale Ni$_{80}$Fe$_{20}$ rings~\citep{Dawidek++2021}. At remanence, each ring in the array typically contained two 180° DWs, which could be driven continuously around the rings' tracks by applying rotating magnetic fields\cite{negoita2012}. Stochastic interactions between DWs at the array's junctions led to both mechanisms for DWs being annihilated from the array and new DW pairs being nucleated, with the balance of these mechanisms depending strongly on the rotating amplitude of the applied field. Thus, the array exhibited a field-dependent emergent response similar to that observed in ASIs. Averaging magnetic behavior over many rings transformed the individual rings' stochastic response into a rich, non-linear, and deterministic aggregate response.

Dawidek \emph{et al.} first used a range of experimental techniques to demonstrate that the ring arrays had the basic physical properties required for reservoir computing. They then used a phenomenological model of their dynamics to demonstrate the classification of digits from the TI-46 database of spoken digits via a time-multiplexed approach, with data being input to the array using the amplitude of a continuously rotating applied field\cite{negoita2012,doi:10.1063/1.4812388} (IOD-1D, DR-D). A recent study by the same team has provided an experimental demonstration of RC with an electrically contacted ring array\cite{vidamourexpt}, where AMR measurements probed the states of the rings.

Interconnected ring arrays have several features that make them highly attractive as reservoirs. Like ASIs, they have numerous geometrical parameters that could tune their dynamic responses. Furthermore, as they consist of many interacting magnetic elements, they offer obvious routes to creating IOD-N reservoirs. However, data input by rotating magnetic fields is unlikely to be energy efficient, and alternative approaches exploiting, e.g., spin-orbit torques, will need to be explored\cite{fukami_2016}.

\section{\label{sec:train}Reservoir Training Methods}
In the previous section, we covered a range of nanomagnetic systems suitable for reservoir computing. Here, we discuss how to train the output layer that receives the reservoir activity to solve various tasks. We present the most popular reservoir training method, known as ridge regression, which requires accumulating all training data and training the reservoir in one step. We also mention a recent technique applicable in an "online learning" setup, where the algorithm progressively adapts its parameters as new data are collected. This new technique enables the reservoir to learn tasks sequentially, which may allow its usage in lifelong learning situations.

Assume that we provide the reservoir with an M-dimensional input signal $\mathbf{s}_i(t)$, where $i$ is an index on the $N_{data}$ different inputs that we can give to the reservoir. Then,  $\mathbf{x}_i(t)$ is an N-dimensional variable that represents measurements in the physical reservoir (or in the traditional neural network setting the activities of the reservoir neurons) as a response to input signal $\mathbf{s}_i(t)$. For each datapoint $i$, we wish to find common parameters ({\it weights}) $\mathbf{W}_{\rm out}$, where $\mathbf{W}_{\rm out}$ is a matrix with dimensions $\rm K \times \rm N$, so that $\mathbf{W}_{\rm out}\mathbf{x}_i(t)=\mathbf{y}_i(t)$, with $\mathbf{y}_i(t)$ being a K-dimensional desirable signal output. We then construct $\mathbf{X}$ and $\tilde{\mathbf{Y}}$ matrices of dimensions $\rm N \times N_{\rm data}$ and $\rm K \times N_{\rm data}$ respectively, obtained through concatenation (across columns) of the measurements (neuron activities) and the desired outputs. We assume $\rm K=1$, meaning we have one output.
To calculate the parameters  $\mathbf{W}_{\rm out}$, we minimize the error function E of the system's output:
\begin{align}
    \rm E=\left(\mathbf{W}_{\rm out} \mathbf{X}-\tilde{\mathbf{Y}}\right) \left(\mathbf{W}_{\rm out} \mathbf{X}-\tilde{\mathbf{Y}}\right)^{\rm T}
    +\beta \mathbf{W}_{\rm out}\mathbf{W}_{\rm out}^{\rm T}
      \label{Eq:Error_ridge}
\end{align}
where $\beta$ is the scaling factor of a term known as the L2 penalty, which penalizes large weights. The method is known as ridge regression and is the most commonly used in the application of reservoir computing.
We can find a closed-form solution to this minimization problem by setting the gradient of $\rm E$ equal to zero: 
\begin{align}
     \mathbf{W}_{\rm out}=\tilde{\mathbf{Y}}\mathbf{X}^{\rm T} \left(\mathbf{X}\mathbf{X}^{\rm T}+\beta \mathbf{I}_{\rm N}\right)^{\rm -1}
    \label{Eq:Ridge}
\end{align}
The solution holds for $\rm K>1$ since we independently minimize every output. While attractive for its simplicity, the ridge regression algorithm is not appropriate for scenarios where the training dataset is not fixed {\it a priori} but increases over time. In particular, robotics applications, reinforcement learning, and {\it lifelong learning} scenarios require algorithms that continuously update their parameters as new data become available. Moreover, solving Eq.~\ref{Eq:Ridge} can be challenging when the matrix to be inverted is very large or rank deficient. 

For this reason, previous research has also adapted iterative learning algorithms to minimize a generic error function, which is not constrained to the mean-squared error. Given an arbitrary cost function $\rm E$, the output weights are optimized iteratively through gradient descent:
\begin{align}
    \mathbf{W}_{\rm out}(n+1)=\mathbf{W}_{\rm out}(n)-\eta \nabla_{\mathbf{W}_{\rm out}} \rm E
\end{align}
where $n$ is the iteration number. 
Alternatively, we can use complex gradient descent methods such as RMSProp or Adam \cite{ruder2016overview, kingma2014adam}, which exploits the first and second-order momentum of the derivatives. Such iterative algorithms are known as {\it online} methods.

 More recently, a sparse online learning algorithm (SpaRCe) has been proposed\cite{Manneschi2021:Sparse}. SpaRCe introduces one threshold per neuron, which is learnable by minimizing the same cost function for the output weights. SpaRCe boosts the performance of online learning in reservoirs applied to classification problems while alleviating the issue of catastrophic forgetting. The latter is a fundamental problem in machine learning; new knowledge overrides older memories when the algorithm learns tasks sequentially. Catastrophic forgetting imposes additional challenges when considering the application of machine learning in {\it lifelong learning} scenarios and is a particularly significant problem for recurrent networks. SpaRCe performs exceptionally well in cases where the reservoir measurements are highly correlated. Since this method doesn't affect the reservoir dynamics, it synergizes well with {\it in materia} reservoirs. Although more time-consuming than the one-step regression, it may enable functionalities that are not possible otherwise, as it improves performance over standard "online methods" in classification problems.

Despite the recent advantages in training methods, and while we consider reservoir computing a promising paradigm for {\it in materia} computing, we do not expect that single reservoirs will be able to compete with more complex structures in general. However, it is possible to achieve competitive performance for specific problems. In a comparative study\cite{Manneschi2021:Sparse} between hierarchical reservoirs and a well-established recurrent network architecture known as Long Sort-Term Memory (LSTM) with the same number of learnable parameters, the reservoirs achieved better performance in the permuted sequential MNIST task. The reservoir learning rule does not need to unravel dependencies in time when finding gradients, reducing the algorithmic complexity by factor T compared to the LSTM, where T is the length of input signals (here 784). These advantages in terms of complexity are expected to translate to reduced energy costs.

\section{Simulation tools}
Many tools are available to model nanoscale magnetic systems, ranging from general-purpose, full-physics simulators to high-level, special-purpose phenomenological models. These tools are essential to developing magnetic RC platforms; experimental demonstrations require challenging device fabrication and subsequent high-throughput characterization of the devices' responses to large quantities of input data. Simulation-based approaches are attractive for scoping functionality when combining these challenges with the wealth of systems and phenomena useful for RC. 

However, simulations of RC also have their challenges. RC requires modeled devices to receive extended streams of input stimuli over timescales at a high computational expense. Furthermore, there is usually a trade-off between the accuracy with which the simulation approach replicates physical phenomena (e.g., magnetization dynamics, the effects of temperature, and materials defects) and their computational cost. We will briefly review the different simulation approaches used to model RC in magnetic materials and discuss where they are best applied.

\subsection{General purpose physical simpulators}
General-purpose physical simulators are powerful modeling software packages that can model a diverse range of nanomagnetic systems. 

{\bf Atomistic Solvers}, such as \textsc{Vampire}\cite{Evans++2014:Vampire}, allow atomic scale simulation of magnetic materials. Magnetic moments are assumed to be localized to atomic sites, and their dynamics are modeled classically via the LLG equation. Modeling materials with this exquisite fidelity allows physically accurate simulations of thermal effects, defects, interfacial interactions, and non-uniform spin textures but at a very high computational cost; it is prohibitively costly to simulate devices with dimensions >100 nm. Consequently, atomistic models are generally poorly suited to exploring RC unless the systems in question are smaller than those we could typically study experimentally\cite{dale2021computing}. 

{\sc \bf Micromagnetic solvers}, such as OOMMF\cite{OOMFuserguide}, NMag\cite{NMag2007}, and {\sc MuMax3}\cite{Vansteenkiste++2014}, model magnetisation as a continuous vector field $\mathbf{M}(\mathbf{r})$, using finite difference or finite element numerical methods. Typically, a model is discretized into individual cells smaller than the exchange length (i.e., the characteristic length scale of a domain wall). Within these cells, we consider magnetization to be uniform. Cells are usually a few nanometres in size and thus represent the magnetic moments of several hundred atoms each. Similar to atomistic solvers, the classical LLG equation models dynamics. Thermal effects may be introduced by including a thermal noise term, resulting in a Langevin thermostat for the system\cite{Leliaert2017}. Since we assume that each cell has a fixed magnetic moment, this approach is limited to temperatures away from the Curie temperature, where we expect large fluctuations in the length of the moment. 

The cells in micromagnetic approaches are typically two orders of magnitude larger than those in atomistic simulations. Therefore, they are substantially less computationally expensive to run. Systems with lateral dimensions \textasciitilde$\mu m$ are easily accessible, especially when using GPU accelerated packages such as {\sc MuMax3}. While these can be used to model RC in modestly sized systems\cite{ababei__2021,Pinna++2020}, the sheer amount of input data required for training can present computational challenges. They are also poorly suited to simulating large systems such as large ASIs or interconnected ring ensembles. Micromagnetic simulations are often best suited to validating the outputs of higher level simulators or training fast, machine learning-based models of system behaviour\cite{chen_2022}.   

\subsection{Special-purpose phenomenological simulators}
The limited applicability of general-purpose simulators to modeling RC stems from the many degrees of freedom they must model. However, simulators specialized to systems of a given class can describe the basic physical behaviors with substantially fewer degrees of freedom.

For example, each island in a typical ASI would consist of \textasciitilde2000 cells with 2 degrees of freedom each if simulated within a micromagnetic framework. At a phenomenological level, it could be represented by a single bistable vector within an Ising model. The GPU-accelerated flatspin simulator\cite{jensen2022flatspin} takes this approach. The simulator has been designed to simulate the dynamics of ASIs as collections of bistable nano-magnets arranged on a lattice, approximated as point dipoles interacting through dipole-dipole coupling. With these approximations, it is possible to model systems comprised of millions of islands. Model predictions were validated against experimental results and other models and allowed simulations demonstrating the applicability of ASIs to RC with modest computational costs\cite{Jensen++2018:spinice}.

RingSim\cite{Dawidek++2021,vidamour2022quantifying}, a simulator designed to predict the behaviors of interconnected nanoring ensembles, takes a similar phenomenological approach. The simulator follows agent-based modeling principles: the active agents are domain walls that are instanced into the model and interact stochastically with a rotating field and other DWs situated in neighboring rings. With this model, it was possible to demonstrate the feasibility of performing RC with a system that would be entirely inaccessible using standard micromagnetic approaches\cite{Dawidek++2021,vidamour2022quantifying}.  

Simple phenomenological models have been used to model a range of other systems, including STOs\cite{furuta_2018}, DW Oscillators \cite{ababei__2021} and super-paramagnet ensembles\cite{Welbourne++2021}. These models are similar in that they sacrifice the detail and accuracy of their descriptions of physics to reduce computational expense. These are appropriate tradeoffs for studies aiming to demonstrate the basic feasibility of RC with a given system as a stepping stone to experimental studies; even predictions from highly detailed atomistic or micromagnetic models are expected to show some variance from real-world devices.

\section{Characterisation beyond benchmark tasks}\label{CHARC}

The suitability of nanomagnetic systems for RC is usually established by performing standard benchmark tasks such as time series prediction or speech recognition (for a review of some key benchmarks, see supplemental information). Evaluating reservoirs in this way provides limited characterization; different tasks require different computational properties. Thus, strong performance in a single task does not guarantee broader usefulness as a reservoir nor scalability to more complex problems.  

In principle, one may achieve a better understanding by measuring task-agnostic reservoir metrics, which characterize a reservoir's computational properties beyond specific benchmarks. Three commonly used metrics are Kernel Rank (KR) \cite{legenstein2007edge}, 
Generalisation Rank (GR) \cite{legenstein2007edge}, 
and Linear Memory Capacity (MC) \cite{Jaeger2002,Dambre2012}.
KR measures the ability of a reservoir to separate different inputs to different reservoir states. GR is the ability of a reservoir to generalize similar inputs to the same reservoir states, and MC is the amount of linear memory within the system. Other metrics have also been proposed\cite{Love_2021}, and careful research will be required to establish which groupings offer the most informative characterizations of a reservoir's computational properties.       

The optimal values of metrics are highly task-dependant. For example, a system with a high GR is susceptible to noisy inputs, whereas a low GR is less sensitive. Depending on the task, these may reflect a desired or undesired property; a noisy input would benefit from a low GR, but a precise and sensitive input would benefit from a high GR. Nonetheless, metrics knowledge can help optimize reservoir design for a specific problem. For instance, if a task requires a particular memory length, knowing which device designs provide the appropriate timescales would lead to a more efficient design process than fabricating several reservoirs and testing them on the specific task.

A step in this direction is CHARC \cite{Dale++2019:PRSA}, a framework for exploring the behavior spaces of families of dynamical systems. 
Traditional search-based methods search for reasonable solutions to a given problem. Instead, CHARC explores the entire behavior space to characterize how well a given set of systems (such as the nanomagnetic systems in this paper) exhibit various dynamical properties usable for solving specific problems.
CHARC defines the space of behaviors by a set of $n$ user-supplied metrics that define an $n$-dimensional \textit{behavior space}. 

It then explores the input parameters to determine the range of behaviors accessible in this space. Using a range is more appropriate for characterizing a system's overall potential than optimizing the parameters for some specific behavior. CHARC uses a novelty search algorithm\cite{Lehman2008,lehman2010efficiently}, an evolutionary algorithm purely explorative, to find sets of input parameters that result in relatively uniformly distributed behaviors over the behavior space. The system is characterized by the volume of behavior space it can access.

CHARC is typically applied to a 3-dimensional behavior space defined by KR, GR, and MC, but it also allows the configuration of alternative measures; there is no claim by the authors of CHARC that these measures are the best for mapping a behavior space \cite{Dale++2019:PRSA}. Given a sufficiently fast and accurate simulator, CHARC can be used to find potentially compelling phenomena to then test in hardware experiments. The results of these experiments can then refine the simulator, creating a closed software improvement loop.

\section{Challenges and outlook}

{\bf Experimental Realisations} Thus far, most studies have only explored nanomagnetic RC in simulation. It is now critical that the most promising proposals are transferred to experimental demonstrations. The challenges here are not a lack of methods to input signals into materials or measure well-established materials' responses but the complexity of the proposed devices and the measurement infrastructure required for proof-of-principle experiments. The latter needs to apply and measure signal trains in substrate-compatible formats at speeds up to GHz. While these challenges are substantial, robust functionality can be demonstrated only via these experimental prototypes under real-world conditions and constraints. While we expect a system computing using material dynamics to be inherently more efficient, such prototypes will alow an accurate measurement of energy consumption\cite{zhong2022memristor} and drive future device improvements.	

{\bf Scalability} Once experiments demonstrate basic functionality, it is essential to examine the scalability of proposed RC systems. For example, for simple IOD-1D, time-multiplexed implementations of RC, it will be essential to examine how computational power is enriched if these devices create IOD-N networks, either via external interconnects or via {\it in materia} interactions. One needs to explore how computational power scales as the size and complexity of systems increase. Computational power will be particularly critical when exploiting {\it in materia} interaction as these will have natural length scales beyond which individual inputs and outputs of a reservoir will not directly interact. Meta-reservoirs, i.e., systems consisting of multiple interconnected reservoirs with different computational properties, should also be explored. Such architectures may likely have substantially greater power than their constituent parts\cite{Manneschi_2021}. In all of these cases, simulations will be an essential tool for exploration. These allow evaluation of the ultimate computational potential of a material system by ignoring the physical confines of interfacing in the first instance.

{\bf Algorithms} The simplicity of the training algorithms RC uses is another critical element for the popularity of RC in the spintronics community. However, this simplicity also has drawbacks; training RC online with the simplest algorithms was challenging until recent methods\cite{Manneschi2021:Sparse} improved its performance by efficiently increasing algorithmic complexity. We pay a small price for improving learning speed and resilience to catastrophic forgetting. Similarly, to achieve {\it Scalability}, we need to optimize the interconnectivity between the reservoirs or their timescales\cite{manneschi2021exploiting}. Typically, however, techniques for finding appropriate parameters require precise mathematical reservoir models, and in spintronic devices, such models may only sometimes be available. Techniques that allow for automated tuning of the parameters of mathematically agnostic reservoirs will be transformative. 

{\bf Evaluation} Task-agnostic metrics offer a powerful platform for understanding the computational properties of potential reservoirs. With the wealth of nanomagnetic systems available for this purpose, careful evaluation of these metrics will be essential for understanding their relative strengths and weaknesses. We do not believe such evaluation will reveal a single system as inherently superior. A wealth of factors must be considered, including power consumption, operating speed, and production cost. More likely, a thorough evaluation of device concepts will reveal what applications they would best suit, whether in lower power edge-computing systems or high-throughput data co-processors, and how nanomagnetic RC systems compare to other competitor technologies. In all cases, it will be essential to recognize the heterotic nature of RC, i.e., conventional electronic systems must augment the reservoir to create input and output layers, all with their constraints and overheads.  

{\bf Applications} So far, reservoir-based spintronic devices have solved simple benchmark problems. While this is inevitable at the earlier stages of research, such toy problems serve only as proof of concept. They are inappropriate for the evaluation of the reservoirs and for attracting a more general interest in the technology. Identifying more challenging tasks within application areas where the spintronics devices may be transformative is necessary. At this stage, it is hard to imagine that spintronic-based RC will serve as general-purpose devices; we expect that there are particular niche areas for which they may be suited. For instance, in the context of edge computing, a promising direction may be that of \emph{smart sensors}, where we would like to offload low-energy preprocessing on the chip. Generally, RC maybe also boost existing methods where additional memory is helpful by adding only a small overhead. For instance, in robotics, the advantages of augmenting existing architectures with a reservoir are demonstrated in the problem of visual place recognition\cite{ozdemir2022echovpr}. For this, interfacing spintronics technology with other hardware may be crucial for the further development of the devices.

\section*{Supplementary Material}
The supplementary material describes the Echo State Network, a fundamental neural network reservoir model, and some typical benchmarks used in reservoir computing.

\begin{acknowledgments}
DAA, TJH, LM, CS, ITV, EV acknowledge funding from the EPSRC MARCH project EP/V006339/1. 
DG, MFKHM, SOK, SS, MAT acknowledge funding from the EPSRC MARCH project 
EP/V006029/1; SOK, SS, MAT also acknowledge partial funding from the EPSRC SpInspired project EP/R032823/1. DAA, TJH, MOAE and EV also acknowledge funding from the EPSRC project EP/S009647/1. DAA, TJH, GV acknowledge Horizon 2020 FET-Open SpinEngine (Agreement no 861618). ITV acknowledges a DTA-funded Ph.D. studentship from EPSRC.
CW acknowledges doctoral funding from the Department of Computer Science, University of York.

\end{acknowledgments}

\bibliography{aipmain}

\end{document}